\documentstyle[eqsecnum,aps,preprint,epsf]{revtex}

\makeatletter
\def\btt#1{{\tt$\backslash$#1}}
\newcommand{\be}{\begin{equation}}
\newcommand{\ee}{\end{equation}}
\newcommand{\bea}{\begin{eqnarray}}
\newcommand{\eea}{\end{eqnarray}}
\newcommand{\bref}[1]{(\ref{#1})}
\makeatother

\begin{document}
\def\btt#1{{\tt$\backslash$#1}}
\draft
\title{{\bf MNS Parameters from Neutrino Oscillations, Single Beta Decay and 
 Double Beta Decay }}
\author{K. MATSUDA, N. TAKEDA, T. FUKUYAMA,}
\address{Department of Physics, 
        Ritsumeikan University, \\
        Kusatsu, Shiga 525-8577, Japan}
\author{and \\ H. NISHIURA}
\address{Department of General Education,
        Junior College of Osaka Institute of Technology, \\
        Asahi-ku, Osaka 535-8585, Japan}

\date{Dec. 21, 2000}

\maketitle



\begin{abstract}
We examine the constraints on the MNS lepton mixing matrix 
from the present and future experimental data of the neutrino oscillation, 
tritium beta decay, 
and neutrinoless double beta decay for Majorana neutrinos.
We show that the small mixing angle solutions for solar neutrino problem are 
disfavored for small averaged mass ($\langle m_\nu \rangle$) 
of neutrinoless double beta decay ($\leq 0.01$ eV) 
in the inverse neutrino mass hierarchy scenario. 
This is the case even in the normal mass hierarchy scenario 
except for very restrictive value of the averaged neutrino mass ($\overline{m_\nu}$) 
of single beta decay.  
The lower mass bound for $\overline{m_\nu}$ 
is given from the present neutrino oscillation data.
We obtain some relations between $\langle m_\nu \rangle$ and $\overline{m_\nu}$. 
The constraints on the Majorana $CP$ violating phases are also given.

\end{abstract}
\pacs{
PACS number(s): 14.60Pq. 23.40.-s 13.10.+q}

\section{Introduction}
The recent neutrino oscillation experiments\cite{skamioka} 
have shown that neutrino do indeed have masses.
On the other hand, 
the experiments intending to determine directly neutrino mass are also on going. 
The upper limit of the averaged neutrino mass $\overline{m_{\nu}}$ defined by
\begin{equation}
\overline{m_{\nu}}^2 \equiv  \sum _{j=1}^{3}|U_{ej}|^2m_j^2
\label{betamass}
\end{equation}
from the tritium $\beta$ decay is 2.2 eV 
\cite{weinheimer}.
Here $U_{a j}$ is the Maki-Nakagawa-Sakata (MNS) left-handed lepton 
mixing matrix which combines the weak eigenstate neutrino 
($a=e,\mu$ and $\tau$) with the mass eigenstate neutrino of 
mass $m_j$ ($j$=1,2 and 3).  
The $U$ takes the following form in the standard representation \cite{fuku}:
\begin{equation}
U=
\left(
\begin{array}{ccc}
c_1c_3&s_1c_3e^{i\beta}&s_3e^{i(\rho-\phi )}\\
(-s_1c_2-c_1s_2s_3e^{i\phi})e^{-i\beta}&
c_1c_2-s_1s_2s_3e^{i\phi}&s_2c_3e^{i(\rho-\beta )}\\
(s_1s_2-c_1c_2s_3e^{i\phi})e^{-i\rho}&
(-c_1s_2-s_1c_2s_3e^{i\phi})e^{-i(\rho-\beta )}&c_2c_3\\
\end{array}
\right).\label{CKM}
\end{equation}
Here $c_j=\cos\theta_j$, $s_j=\sin\theta_j$ 
($\theta_1=\theta_{12},~\theta_2=\theta_{23},~\theta_3=\theta_{31}$). 
Note that three 
$CP$ violating phases, $\beta$ , $\rho$ and $\phi$ appear in $U$ for 
Majorana particles \cite{bilenky}. 
On the other hand, the most recent experimental upper bound for the averaged 
mass $\langle m_{\nu} \rangle$  defined by 
\begin{equation}
\langle m_{\nu} \rangle\equiv |\sum _{j=1}^{3}U_{ej}^2m_j| 
\label{betabetamass}
\end{equation}
for Majorana neutrinos from the neutrinoless double beta decay 
($(\beta\beta)_{0\nu}$) is given by 
$\langle m_{\nu} \rangle<0.2$ eV in the absence of right-handed lepton current 
\cite{baudis}. 
The next generation experiments such as GENIUS\cite{genius}, CUORE\cite{cuore}, 
MOON\cite{moon} are anticipated to reach a considerably more stringent limit 
$\langle m_{\nu} \rangle<0.01-0.001$ eV.
In these situations it is necessary to confine the MNS parameters 
and the Majorana phases by incorporating all these direct and indirect experiments.
To disentangle the complicated correlations among the data from many different 
kinds of experiments, we proposed a graphical method \cite{matsu}\cite{matsu2}.
This method enables us to grasp the geometrical relations among 
the parameters of masses, mixing angles, $CP$ phases etc. 
and obtain the constraints on them more easily than the analytical calculations
\cite{fuku}\cite{fuku2}\cite{nishi}. 
In this work we apply this method 
to obtain the constraints on the MNS parameters by using 
the neutrino oscillation, 
$(\beta\beta)_{0\nu}$ and single beta decay experiments.
\par
This article is organized as follows. In section 2 we review the 
graphical representations of the complex masses and the $CP$ violating phases 
appeared in $\langle m_{\nu} \rangle$.  
In section 3 we discuss the constraints among mixing angles, 
$\langle m_\nu \rangle$ and \(\overline{m_\nu}\) from $(\beta\beta)_{0\nu}$, 
single beta decay and neutrino oscillation. 
In section 4 we give the allowed region in the 
$\langle m_{\nu} \rangle$-$\overline{m_\nu}$ plane, 
first irrespective and second respective of the mixing angles.
Constraints on the $CP$ violating phases are discussed in section 5. 
Section 6 is devoted to summary.

\section{Graphical representations of the complex masses 
and the $CP$ violating phases appearing in Neutrinoless double beta decay}
\begin{figure}[htbp]
\centerline{\epsfbox{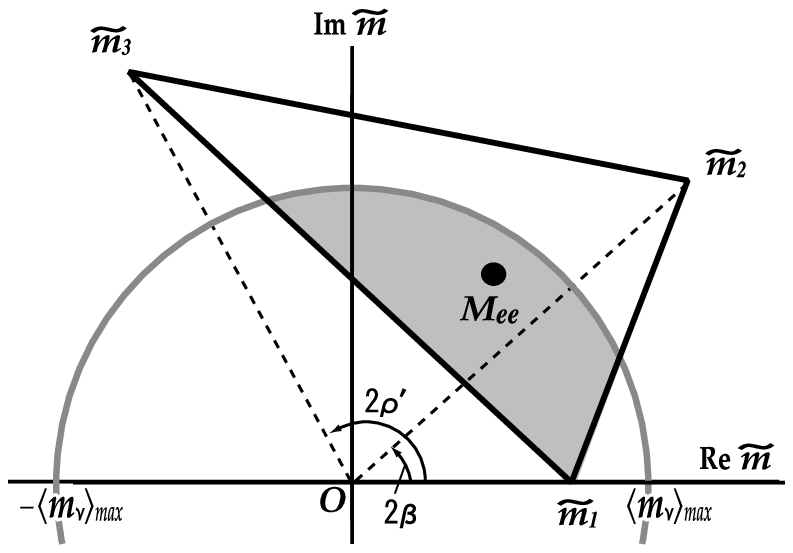}}
\caption{%
Graphical representations of the complex mass \(M_{ee}\) 
and $CP$ violating phases.
The position of $M_{ee}$ 
is within the triangle formed by the three points $\widetilde{m_i} (i=1,
2\mbox{ and }3)$ which are defined in Eq.(2.3).
The allowed position of \(M_{ee}\) is in the intersection (shaded area) 
of the inside of this triangle and the inside of the circle of radius 
\(\langle m_\nu \rangle_{max}\) around the origin.}
\label{fig1}
\end{figure}

\begin{figure}[htbp]
\centerline{\epsfbox{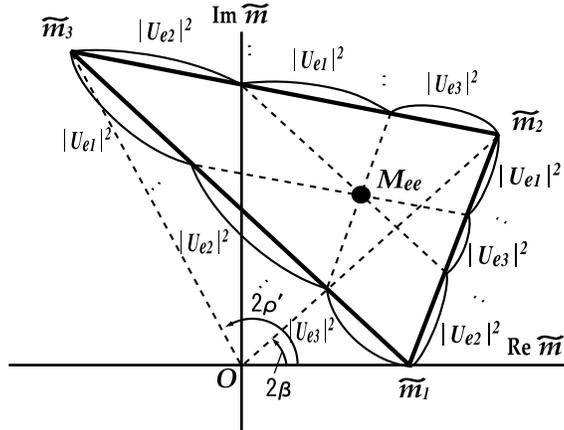}}
\caption{%
The relations between the position $M_{ee}$ and $U_{ei} (i=1,2\mbox{ and }3)$ 
components of MNS mixing matrix. 
The three mixing elements $|U_{ej}|^2(j=1,2\mbox{ and }3)$ 
indicate the division ratios for the three portions of 
each side of the triangle which are divided by the parallel 
lines to the side lines of the triangle passing through $M_{ee}$.}
\label{fig2}
\end{figure}

Let us review our graphical representations of the complex mass and 
the $CP$ violating phases appeared in $(\beta\beta)_{0\nu},$ 
which are proposed in \cite{matsu}\cite{matsu2}. 
The averaged mass $\langle m_{\nu} \rangle$ of 
Eq.(\ref{betabetamass})\cite{doi} is the absolute value of averaged 
complex masses for Majorana neutrinos,  
\begin{eqnarray}
\langle m_{\nu} \rangle  & =& |M_{ee}|,\label{eq761}
\end{eqnarray}
where the averaged complex mass $M_{ee}$ is, after suitable phase convention, 
defined by
\be
M_{ee} \equiv \sum _{j=1}^{3}U_{ej}^2m_j 
       \equiv \sum _{j=1}^{3}|U_{ej}|^2\widetilde{m_j}. 
\label{cmass}
\ee
Here we have defined the complex masses $\widetilde{m_i} (i=1,2\mbox{ and }3)$ by
\begin{eqnarray}
\widetilde{m_1} & \equiv& m_1, \qquad
\widetilde{m_2}  \equiv e^{2i\beta }m_2, \nonumber\\
\widetilde{m_3} & \equiv&  e^{2i( \rho -\phi )}m_3\equiv e^{2i \rho '}m_3.
\label{tildemass}
\end{eqnarray}
Since \(\sum^3_{j=1} |U_{ej}|^2=1\), the position of $M_{ee}$ in a complex 
mass plane is within the triangle formed by the three vertices 
$\widetilde{m_i} (i=1,2\mbox{ and }3)$ if the magnitudes of $|U_{e j}|^2  
(j=1,2\mbox{ and }3)$ 
are unknown (Fig.\ref{fig1}). 
This triangle is referred to the complex-mass triangle\cite{matsu}
\cite{matsu2}. 
The three mixing matrix elements $|U_{e j}|^2  (j=1,2\mbox{ and }3)$ indicate 
the division ratios for the three portions of each side of the triangle 
which are divided by the parallel lines to the side lines of the triangle 
passing through the $M_{ee}$ (Fig.\ref{fig2}). 
The $CP$ violating phases $2\beta$ and $2\rho^\prime$ represent 
the rotation angles of 
$\widetilde{m_2}$ and $\widetilde{m_3}$ around the origin, respectively. 
Since $\langle m_{\nu} \rangle=|M_{ee}|,$ 
the present experimental upper bound on $\langle m_{\nu} \rangle$ 
(we denote it $\langle m_{\nu} \rangle_{max}$) 
indicates the maximal distance of the point $M_{ee}$ from the origin 
and forms the circle in the complex plane (Fig.\ref{fig1}). 
\section{Constraints on mixing angles from neutrino oscillations, 
neutrinoless double beta and single beta decays}
\begin{figure}[htbp]
\centerline{\epsfbox{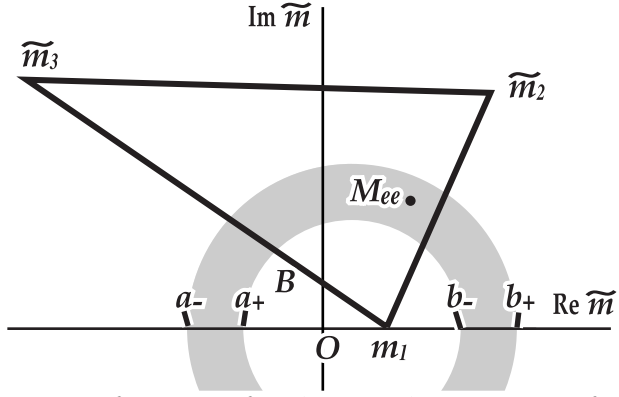}}
\caption{%
The allowed region of position for the complex mass $M_{ee}$ 
from $(\beta\beta)_{0\nu}$ obtained from Eq.(3.2). In this figure, 
\(a_{-} \equiv\) \(|U_{e1}|^2 m_1-\) \((|U_{e2}|^2 m_2\) \(+|U_{e3}|^2 m_3)\),
\(a_{+} \equiv\) \(|U_{e1}|^2 m_1-\) \(||U_{e2}|^2 m_2\) \(-|U_{e3}|^2 m_3|\),
\(b_{-} \equiv\) \(|U_{e1}|^2 m_1+\) \(||U_{e2}|^2 m_2\) \(-|U_{e3}|^2 m_3|\) and 
\(b_{+} \equiv\) \(|U_{e1}|^2 m_1+\) \((|U_{e2}|^2 m_2\) \(+|U_{e3}|^2 m_3)\).}

\label{fig3}
\end{figure}

We first discuss the constraints on the mixing angles from 
$(\beta\beta)_{0\nu}$ and neutrino oscillations.
The averaged mass $\langle m_{\nu} \rangle$ obtained from 
$(\beta\beta)_{0\nu}$ is given by the absolute value of averaged complex mass 
$M_{ee}$ as in Eq.\bref{cmass}.
The $CP$ phases $\beta$ and $\rho '$ in $M_{ee}$ may have still time to be 
determined. 
Let us move these two parameters freely and find constraints irrespective of 
these phases.
It goes from Eq.\bref{cmass} that 
\be
|M_{ee}-|U_{e1}|^2\widetilde{m_1}| =
 ||U_{e2}|^2\widetilde{m_2}+|U_{e3}|^2\widetilde{m_3}|. 
\ee 
Hence one finds
\be
 ||U_{e2}|^2m_2-|U_{e3}|^2m_3| \le|M_{ee}-|U_{e1}|^2m_1| \le |U_{e2}|^2m_2
 +|U_{e3}|^2m_3. 
\label{dounats}
\ee  
Therefore, the position of $M_{ee}$ in a complex mass plane is 
within the annulus shown in Fig.\ref{fig3}.
As we mentioned, \(M_{ee}\) must also be inside of the circle with radius 
\(\langle m_\nu \rangle_{max}\). 
\par 
So we obtain the consistency condition
\bea
&-\langle m_{\nu} \rangle_{max} < |U_{e1}|^2m_1-||U_{e2}|^2m_2-|U_{e3}|^2m_3| 
\equiv a_+& 
  \nonumber \\
&\text{and}& \nonumber\\
&a_-\equiv |U_{e1}|^2m_1-|U_{e2}|^2m_2-|U_{e3}|^2m_3 < 
\langle m_{\nu} \rangle_{max}.&
\label{mixingconstraints}
\eea 
\par
In order to obtain the constraints on the parameters from the consistency 
condition\bref{mixingconstraints} 
let us use, instead of \(m_1\), \(m_2\) and \(m_3\),
$\overline{m_{\nu}}$, $\Delta m_{12}^2\equiv m_2^2-m_1^2$ and  
$\Delta m_{23}^2\equiv m_3^2-m_2^2$ which are measurable in single beta decay or 
neutrino oscillation experiments. 
Namely, by inserting the relations  
\bea
m_2 & =&\sqrt{m_1^2+\Delta m_{12}^2}, \label{eq1220-01}\\
m_3 & =&\sqrt{m_2^2+\Delta m_{23}^2}
      =\sqrt{m_1^2+\Delta m_{12}^2+\Delta m_{23}^2} \label{eq1220-02}
\eea
into Eq.\bref{betamass} with the unitarity condition that 
$|U_{e2}|^2=1-|U_{e1}|^2-|U_{e3}|^2$, 
we obtain the following expressions for $m_1$, $m_2$, and $m_3$ 
\cite{ichep}:
\bea
m_1 & =&
  \sqrt{\overline{m_{\nu}}^2-(1-|U_{e1}|^2)\Delta m_{12}^2-|U_{e3}|^2
  \Delta m_{23}^2}, \label{eq1220-11}\\
m_2 & =&\sqrt{\overline{m_{\nu}}^2+|U_{e1}|^2\Delta m_{12}^2-|U_{e3}|^2
  \Delta m_{23}^2},\\
m_3 & =&
  \sqrt{\overline{m_{\nu}}^2+|U_{e1}|^2\Delta m_{12}^2+(1-|U_{e3}|^2)
  \Delta m_{23}^2}.\label{eq1220-12}
\eea
\par
It should be noted from $0 \le m_1^2$ that the neutrino mass 
$\overline{m_{\nu}}$ in Eq.(1.1) is predicted to have 
lower bound as
\be
\sqrt{(1-|U_{e1}|^2)\Delta m_{12}^2+|U_{e3}|^2\Delta m_{23}^2} \le 
\overline{m_{\nu}}. 
\label{eq122001}
\ee
\par
Now we discuss constraints among MNS mixing parameters 
$|U_{e1}|^2$, $|U_{e3}|^2$, averaged neutrino masses $\overline{m_{\nu}}$, and 
$\langle m_{\nu} \rangle$ from the consistency condition 
\bref{mixingconstraints} by using the data obtained from 
the neutrino oscillation experiments. 
In the following discussions we consider two scenarios 
for neutrino mass hierarchy:
\par
(A) normal mass hierarchy where $m_1 \sim m_2 \ll m_3$ and 
\par
(B) inverse mass hierarchy where $m_1 \ll m_2 \sim m_3.$\\
For the case (A), we have 
\be
\Delta m_{23}^2 =\Delta m_{atm}^2\cong 0.0032 ~\text{eV$^2$ }
\label{eq111501}
\ee
from the Super-Kamiokande atmospheric neutrino oscillation experiment
\cite{skamioka}. For the solar neutrino deficit \cite{solar} we have 
\be
\Delta m_{12(MSW)}^2 =\Delta m_{solar(MSW)}^2\cong 0.00001 ~\text{eV$^2$ }
\ee
for  MSW solutions and  
\be
\Delta m_{12(Just~So)}^2 =\Delta m_{solar(Just~So)}^2
\cong 4.3\times 10^{-10} ~\text{eV$^2$ }
\ee
for vacuum oscillation (Just So) solution. 
From the recent CHOOZ reactor experiment\cite{chooz} we have 
\be
|U_{e3}|^2<0.03. \label{eq111502}
\ee
So two averaged masses  $\overline{m_\nu},~\langle m_{\nu} \rangle$ and
 $|U_{e1}|^2$ are left as free parameters. 
The \(|U_{ei}|^2\) runs over the large mixing angle (LMA-MSW) 
and the small mixing angle (SMA-MSW) solutions 
to the solar neutrino problem. 

For the case (B), we may simply exchange the suffix \(1\) for \(3\) 
in the arguments of normal mass hierarchy.
That is, Eqs.(\ref{eq111501})-(\ref{eq111502}) are replaced by
\begin{eqnarray}
&&\Delta m_{12}^2 =\Delta m_{atm}^2\cong 0.0032 ~\text{eV$^2$ }, 
\label{eq121901}\\
&&\Delta m_{23(MSW)}^2 =\Delta m_{solar(MSW)}^2\cong 0.00001 
~\text{eV$^2$ }, \\
&&\Delta m_{23(Just~So)}^2 =\Delta m_{solar(Just~So)}^2
\cong 4.3\times 10^{-10} ~\text{eV$^2$ },\\
&&|U_{e1}|^2<0.03,\label{eq121902}
\end{eqnarray}
respectively.  So in this case $|U_{e3}|^2$ in place of $|U_{e1}|^2$ together 
with two averaged masses are left unknown.\\ 
\par

Substituting the observed values in oscillation experiments into 
Eq.\bref{mixingconstraints}, we obtain the constraints on the two averaged masses 
as well as $|U_{e1}|^2$ for the normal mass hierarchy and  $|U_{e3}|^2$ 
for the inverse mass hierarchy (Fig.\ref{fig4}). 
Here we have set the typical values for 
$\langle m_{\nu} \rangle_{max}.$
So the allowed regions are given in the $\overline{m_{\nu}}$-$|U_{e1}|^2$ 
plane for fixed values of $\langle m_{\nu} \rangle_{max}$, 
$\Delta m_{solar}^2$, $\Delta m_{atm}^2$, and $|U_{e3}|^2$ in the normal mass 
hierarchy scenario (Eqs.(\ref{eq111501})-(\ref{eq111502})), and are given in the
$\overline{m_{\nu}}$-$|U_{e3}|^2$ plane for fixed values of 
$\langle m_{\nu} \rangle_{max}$, $\Delta m_{solar}^2$, $\Delta m_{atm}^2$, 
and $|U_{e1}|^2$ in the inverse mass hierarchy scenario 
(Eqs.(\ref{eq121901})-(\ref{eq121902})). 
It is seen from Fig.\ref{fig4} that, for small $\langle m_{\nu} \rangle_{max}$, 
the SMA-MSW  solution to the solar neutrino problems is disfavored for both 
hierarchies.
That is, the SMA-MSW is allowed only in the very narrow region; in 
the normal mass hierarchy $\overline{m_{\nu}} \sim 1\times 10^{-2}$ eV for 
$\langle m_{\nu} \rangle_{max}<0.01$ eV and in the inverse mass hierarchy 
$\overline{m_{\nu}} \sim 1\times 10^{-1}$ eV for 
$\langle m_{\nu} \rangle_{max}<0.1$ eV. 
If $\langle m_{\nu} \rangle_{max}<0.01$ eV, the SMA-MSW in the inverse mass 
hierarchy is excluded irrespective of $\overline{m_{\nu}}$. These values of 
$\langle m_{\nu} \rangle_{max}$ are very marginal to the present experimental 
values and, therefore, the future experiments such as GENIUS\cite{genius}, 
CUORE\cite{cuore}, MOON\cite{moon} will play important roles for choosing 
the neutrino mass hierarchy scenarios. 
Klapdor et al. have discussed about the same subjects
in excellent clearness \cite{Klapdor}. 
We have developed the graphical method, incorporated 
$\overline{m_{\nu}}$ in the analyses and obtained more stringent constraints 
than those in \cite{Klapdor}.  
As for $\overline{m_{\nu}}$, it seems very hard to improve 
the upper limit beyond the present 3eV to 0.5eV \cite {weinheimer}.  
It needs the improvement of the intensity of source or 
the acceptance of spectrometer by factor 100-1000.  
Also the problem of $\overline{m_{\nu}}^2<0$ still remains unsolved \cite{ohshima}.  
Anyhow, however, 
if non zero $\overline{m_{\nu}}$ is measured in the neighborhood 
of the present upper limit, 
then it immediately leads to the LMA-MSW or Just So solution 
and the impacts to the other branches are very large. 
Theoretical upper bound of neutrino masses, for instance, 
from the matching condition for fluctuation powers at the COBE scale and
at the cluster scale for generous parameter space is 0.9eV \cite{fukugita}. 
The Fig.\ref{fig4} also shows 
that $\overline{m_{\nu}}$ is larger than $O(10^{-2})$ eV
for the normal neutrino mass hierarchy scenario and than $O(10^{-1})$ eV 
for the inverse neutrino mass hierarchy scenario irrespective of 
$\langle m_{\nu} \rangle_{max}$, which comes from Eq.(\ref{eq122001}). 
If the GENIUS I and II can not observe the $(\beta\beta)_{0\nu}$ 
for \(\langle m_\nu \rangle \agt 0.01\)eV 
and if LMA-MSW and Just So solutions become disfavored,
then the normal neutrino mass hierarchy scenario 
(each lower three panels of (a) and (b) in Fig.4) can survive.

\section{Constraints on averaged neutrino masses defined 
in neutrinoless double beta decay and in single beta decay}
\par
We will show how the data of $\overline{m_{\nu}}$ give the restriction 
on the complex mass triangle of the $(\beta\beta)_{0\nu}$.
By using Eq.\bref{betamass} and the unitarity condition of $U_{ei}$ , 
one finds that among three mixing matrix elements 
$|U_{ej}|(j=1,2\mbox{ and }3)$ 
only one matrix matrix element is free parameter 
and the others are expressed in terms of it once neutrino masses 
$m_i$ and $\overline{m_{\nu}}$ are fixed:
\bea
|U_{e1}|^2&=&\frac{m_3^2-\overline{m_{\nu}}^2-|U_{e2}|^2(m_3^2-m_2^2)}
                  {m_3^2-m_1^2}, 
\nonumber\\
|U_{e3}|^2&=&\frac{\overline{m_{\nu}}^2-m_1^2-|U_{e2}|^2(m_2^2-m_1^2)}
                  {m_3^2-m_1^2}.
\label{unitary}
\eea
Let us assume that the averaged mass 
in tritium beta decay has the definite value \(\overline{m_{\nu}}\).
Then it goes from Eqs.\bref{betamass}, \bref{unitary} and Fig.\ref{fig2} 
that the position of $M_{ee}$ is restricted to be on the line segment 
$AB$ in Fig.\ref{fig5}(a) for the case 
$\overline{m_{\nu}}<m_2$ or $A^\prime B$ in Fig.\ref{fig5}(b)
for the case $\overline{m_{\nu}}>m_2.$ 
Here from the condition \(0\le |U_{ei}|^2\le 1\) \((i=1,2\mbox{ and }3)\), 
the mixing matrix elements $|U_{ei}|$ range over
\bea
\frac{m_2^2-\overline{m_{\nu}}^2}{m_2^2-m_1^2} 
 & \le&|U_{e1}|^2 \le \frac{m_3^2-\overline{m_{\nu}}^2}{m_3^2-m_1^2}, 
 \nonumber\\
\frac{\overline{m_{\nu}}^2-m_1^2}{m_2^2-m_1^2}  & \ge&|U_{e2}|^2\ge0, 
 \nonumber\\
0 & \le& |U_{e3}|^2 \le \frac{\overline{m_{\nu}}^2-m_1^2}{m_3^2-m_1^2}
\label{u-bound1} 
\eea
for the case $\overline{m_{\nu}}<m_2$ and
\bea
0  & \le&|U_{e1}|^2 \le \frac{m_3^2-\overline{m_{\nu}}^2}{m_3^2-m_1^2}, 
\nonumber\\
\frac{m_3^2-\overline{m_{\nu}}^2}{m_3^2-m_2^2}  & \ge&|U_{e2}|^2\ge0, 
\nonumber\\
\frac{\overline{m_{\nu}}^2-m_2^2}{m_3^2-m_2^2}  & \le& 
|U_{e3}|^2 \le \frac{\overline{m_{\nu}}^2-m_1^2}{m_3^2-m_1^2} 
\label{u-bound2} 
\eea
for the case $\overline{m_{\nu}}>m_2$. 
Eq.\bref{unitary} combined with the analysis of 
the $(\beta\beta)_{0\nu}$ gives additional constraints on the position of 
the complex mass $M_{ee}$ as follows.
\par
The $M_{ee}$ runs over the line segments with free parameter $U_{e2}$ for fixed 
$\beta, ~\rho'.$ (Observed data of $U_{ei}$ will be incorporated later in this section.) 
Here $A$ divides the line $\overline{m_1\widetilde{m_2}}$ by 
$Am_1 :A\widetilde{m_2}=r:s$.
The \(r\) and \(s\) are defined by
\be
r=\frac{\overline{m_{\nu}}^2-m_1^2}{m_2^2-m_1^2},~~ 
s=\frac{m_2^2-\overline{m_{\nu}}^2}{m_2^2-m_1^2}
\quad (r+s=1).
\label{rs}
\ee
Therefore
\be
OA=rm_2e^{2i\beta}+sm_1.
\label{oa}\ee
Hence if we move $\beta$ freely, the position \(A\) occupies a circle 
(dotted circle crossing at $A_{\pm}$ with horizontal axis (Fig.\ref{fig6}(a)).
Likewise, $B$ divides the line $\overline{m_1\widetilde{m_3}}$ by 
$Bm_1:B\widetilde{m_3}=q:p,$
where \(p\) and \(q\) are defined by
\be
p=\frac{m_3^2-\overline{m_{\nu}}^2}{m_3^2-m_1^2},~~ 
q= \frac{\overline{m_{\nu}}^2-m_1^2}{m_3^2-m_1^2} 
\quad (p+q=1).
\label{pq}
\ee
So $OB$ is given by
\be
OB=qm_3e^{2i\rho '}+pm_1.
\label{OB}
\ee
Then if we move $\rho'$ freely, the position \(B\) occupies a circle 
(dotted circle crossing at $B_{\pm}$ with horizontal axis (Fig.\ref{fig6}(a))).
Thus making both $\beta$ and $\rho'$ run freely, $M_{ee}$ on the line segments 
is inside an larger circle, the region bounded by dotted circles passing 
$A_-AA_+$ (Fig.\ref{fig6}). The $A_{\pm}$ and $B_{\pm}$ are easily obtained graphically.
For instance, $A_-$ divides $\overline{(-m_2)m_1}$ as $\overline{(-m_2)A_-}:
\overline{A_-m_1}=\overline{\widetilde{m_2}A}:\overline{Am_1}=
s:r,$ 
where use has been made of the ratio in Fig.\ref{fig5}. 
The other points are also obtained analogously.  
Thus we easily obtain graphically
\be
Max\{\overline{OA_{-}},0\}=
Max\{\frac{m_1m_2-\overline{m_{\nu}}^2}{m_2-m_1},0\} \le 
\langle m_{\nu} \rangle 
\le \overline{OA_{+}}= \frac{m_1m_2+\overline{m_{\nu}}^2}{m_2+m_1}
\label{maxmin1}
\ee
for the case  $\overline{m_{\nu}}\leq m_2.$ 
The $Max\{a,b\}$ indicates the larger value between $a$ and $b.$ 
As for the case of $\overline{m_{\nu}}>m_2$ (Fig.\ref{fig6}(b)), we repeat 
the almost same arguments and obtain
\be
Max\{\overline{OA_{-}},0\}=
Max\{\frac{\overline{m_{\nu}}^2-m_3m_2}{m_3-m_2},0\} \le 
\langle m_{\nu} \rangle \le 
\overline{OA_{+}}= 
\frac{m_3m_2+\overline{m_{\nu}}^2}{m_3+m_2}.
\label{maxmin2}
\ee
Note that Eq.\bref{maxmin2} is 
obtained from Eq.\bref{maxmin1} by exchanging the suffix 3 for 1. 
Summing up the results of Eqs.(\ref{maxmin1}) and (\ref{maxmin2}), 
we obtain the allowed region 
in the \(\langle m_\nu \rangle\)-\(\overline{m_\nu}\) plane(Fig.7). 
In this Figure we have not specified the points on the segments \(AB\) or 
\(A^\prime B\) which means 
that the information of mixing angles are smeared out. 
The $m_1$ is only free parameter and the other masses are expressed in terms of 
$\Delta m_{12}^2$ and $\Delta m_{23}^2$ 
from Eqs.(\ref{eq1220-01}) and (\ref{eq1220-02}).  
If we incorporate the information of the mixing angles (some restricted regions), 
the points on the segments \(AB\) or \(A^\prime B\) in Fig.6 are not specified 
but restricted in some small region and we have more stringent constraints 
in the $\langle m_{\nu} \rangle - \overline{m_{\nu}}$ plane than in Fig.7. 
We list only the cases of LMA-MSW and SMA-MSW  solutions 
in the normal mass hierarchy (Fig.8), 
where we have adopted $|U_{e3}|^2<0.03$, and  $0.3<|U_{e2}|^2<0.7$ for LMA-MSW  
and $1\times 10^{-3}<|U_{e2}|^2<1\times 10^{-2}$ for SMA-MSW .  
The first, second and third row panels correspond to the case 
$m_1=10^{-5}$eV, $10^{-3}$eV, and $10^{-1}$eV, respectively. 
Namely within these conditions of \(|U_{ei}|^2\), 
all masses are fixed and \(CP\) phases are free parameters in each panel.
From Fig.8(f), if GENIUS I gives a new upper bound on $\langle m_{\nu} \rangle$, 
SMA-MSW  with $m_1\sim 10^{-1}$eV is excluded.  
This figure is rather useful for checking models which predict mass spectrum.

\section{Constraints on $CP$ violating phases from 
neutrinoless double beta and single beta decays}
Now we discuss possible constraints on $CP$ violating phases 
\cite{matsu} \cite{matsu2} \cite{minakata} from 
$(\beta\beta)_{0\nu}$ and single beta decay experiments. 
First, we consider the constraints for the typical values of the mixing matrix 
element such as (i) $|U_{e3}|^2 =0$, minimum value of $|U_{e3}|^2,$ and 
(ii) $|U_{e3}|^2 =\frac{\overline{m_{\nu}}^2-m_1^2}{m_3^2-m_1^2},$ 
maximum value of $|U_{e3}|^2,$ in the case where $\overline{m_{\nu}}<m_2.$ 
That is, if $m_i(i=1,2\mbox{ and }3),\overline{m_{\nu}},$ and $U_{e3}$ are given, then 
$U_{ej}(j=1,2)$ are also given from Eq.\bref{unitary}.  
Hence if we impose the observed 
$\langle m_{\nu} \rangle_{max}$, then we can constrain 
the $CP$ violating phases.
\par
For the case $|U_{e3}|^2 =0$, the position of complex mass $M_{ee}$ is 
restricted to be at point \(A\) in Fig.\ref{fig5}(a) as was discussed in section 4. 
Therefore, when we change the $CP$ violating phase $\beta$, $M_{ee}$ moves 
along a circle passing $A_-AA_+$ (Fig.6)
Since this circle must be inside of the circle with the radius 
$\langle m_{\nu} \rangle_{max}$. 
Two these circles intersect at $\beta$ satisfying 
$\overline{OA}$ of Eq.\bref{oa}$=\langle m_{\nu} \rangle_{max}$, that is,
\be
\cos^{-1}\frac{\langle m_{\nu} \rangle_{max}^2-(m_1r)^2-(m_2s)^2}{2m_1m_2rs} 
\le |2\beta| \le \pi \label{eq1220-13}
\ee 
for the case  $\frac{m_1m_2-\overline{m_{\nu}}^2}{m_2-m_1} \le \langle m_{\nu} 
\rangle_{max} \le \frac{m_1m_2+\overline{m_{\nu}}^2}{m_2+m_1}$, 
where \(r\) and \(s\) are defined in Eq.\bref{rs}. 
We have no constraint on $\beta$  for the case 
$\frac{m_1m_2+\overline{m_{\nu}}^2}{m_2+m_1} \le 
\langle m_{\nu} \rangle_{max}$ and  the case $\langle m_{\nu} \rangle_{max}
\le-\frac{m_1m_2-\overline{m_{\nu}}^2}{m_2-m_1}$ is excluded.
\par
For the case $|U_{e3}|^2=\frac{\overline{m_{\nu}}^2-m_1^2}{m_3^2-m_1^2}$, the 
position of complex mass $M_{ee}$ is restricted to be at point \(B\) 
in Fig.\ref{fig5}(a). 
Therefore, when we change the $CP$ violating phase $\rho'$, $M_{ee}$ moves 
along a circle $B_-BB_+$ (Fig.6)
irrespective of 
the $CP$ violating phase $\beta.$ 
Thus we obtain the allowed bounds on 
the $CP$ violating phase $\rho',$ 
\be
\cos^{-1}\frac{\langle m_{\nu} \rangle_{max}^2-(m_1p)^2-(m_3q)^2}{2m_1m_3pq} 
\le |2\rho'| \le \pi
\ee 
for the case $\frac{m_1m_3-\overline{m_{\nu}}^2}{m_3-m_1} 
\le \langle m_{\nu} \rangle_{max} \le 
\frac{m_1m_3+\overline{m_{\nu}}^2}{m_3+m_1}$, 
where $p$ and $q$ are defined in Eq.\bref{pq}.
We have no constraint on $\rho'$ for the case 
$ \frac{m_1m_3+\overline{m_{\nu}}^2}{m_3+m_1} \le 
  \langle m_{\nu} \rangle_{max}$ 
and the case $\langle m_{\nu} \rangle_{max}
\le-\frac{m_1m_3-\overline{m_{\nu}}^2}{m_3-m_1}$ is excluded.
The case for \(\overline{m_{\nu}}>m_2\) can be discussed in the same way.

So far we have considered the special limit of 
mixing angles and smeared one \(CP\) phases. 
In what follows, 
we consider the constraints on both two CP phases by 
incorporating mixing angles by use of Fig.2 and Fig.5.
We obtain the allowed region among \(\overline{m_{\nu}}\), 
\(CP\) violating phases \(\beta\) and \(\rho'\)
for the LMA-MSW solution in the normal and inverse mass hierarchy scenarios.
It is given in Fig.9 and Fig.10 in the case where the \(e\)-\(\mu\) mixing 
is nearly maximal, namely \(|U_{e1}|^2\simeq|U_{e2}|^2 \simeq 1/2\).
In these figures, let us consider the situation 
that  $\langle m_\nu \rangle$ and $\overline{m_{\nu}}$ have the nonzero values 
in the neighborhood of the present or near future upper limits. 
For the case of \(|U_{e3}|^2 \ll O(10^{-2})\), 
one obtains the restriction on \(\beta\) only by any experiments 
because Eq.(\ref{eq1220-13}) is irrespective of \(\rho'\).
However, if \(|U_{e3}|^2\) has a slight non-zero value (\(|U_{e3}|^2=O(10^{-2})\)) 
and if GENIUS II first finds a nonzero $\langle m_\nu \rangle$, 
then one obtains the restrictions on not only \(\beta\) but also \(\rho'\)
((b) and (d) in Fig.9 and Fig.10). 
However \(\beta\) and \(\rho'\) are very sensitive to 
the value in the neighborhood of \(|U_{e2}|^2\simeq 1/2\).
Fig.9 is the case for \(|U_{e2}|^2=0.5\).
In this case, \(\sin\rho'\ll 1\) and \(\sin\beta \simeq 1\) both in 
the normal (Fig.9 (b)) and inverse (Fig.9 (d)) hierarchy scenarios.
The case for \(|U_{e2}|^2 =0.47\) is given in Fig.10.
In this case, \(\sin \beta\simeq 1\) and \(\sin\rho' \simeq 1\) 
in the normal hierarchy (Fig.10 (b))
and \(\sin \beta \ll 1 \) and \(\sin\rho' \simeq 1\) 
in the inverse hierarchy (Fig.10 (d)). 
In these figures, nevertheless, 
the restrictions of \(\beta\) in the normal hierarchy (\(\sin\beta\simeq1\))
and of \(|\rho'-\beta|\) in the inverse hierarchy (\(\sin|\beta-\rho'|\simeq 1\)) 
are insensitive to \(|U_{e2}|^2 \simeq 1/2\).


\section{summary}
In addition to the neutrino oscillation experiments, 
the experiments intending to determine directly neutrino mass, 
such as neutrinoless double beta decay experiments (GENIUS, CUORE, MOON etc.), 
and the tritium beta decay experiments (Mainz, Troitsk etc.) are 
on going and anticipated to reach a considerably more stringent limit. 
In these situations, it is more and more important to consider MNS parameters 
from various experiments and phenomena. 
In this paper, assuming that neutrinos are Majorana particles of three generations, 
we have examined the constraints on the mixing angles 
and Majorana $CP$ violating phases of the MNS lepton mixing matrix 
by analyzing $(\beta\beta)_{0\nu}$ together with   
the neutrino oscillations and the single beta decay.
In this analysis we have seriously considered the $CP$ violating phases appeared 
in $(\beta\beta)_{0\nu}$ and applied the graphical method proposed 
in the previous papers (Fig.1 and 2).
The meaning of incorporating \(CP\) phases is twofold.  
One is, of course, to predict their magnitudes. 
The other is to extract the results which are valid irrespective 
of the magnitudes of \(CP\) phases. 
Sections 3 and 4 correspond to the latter case 
and section 5 to the former case. 

Our results are in order:
\par
\begin{figure}[htbp]
\centerline{\epsfbox{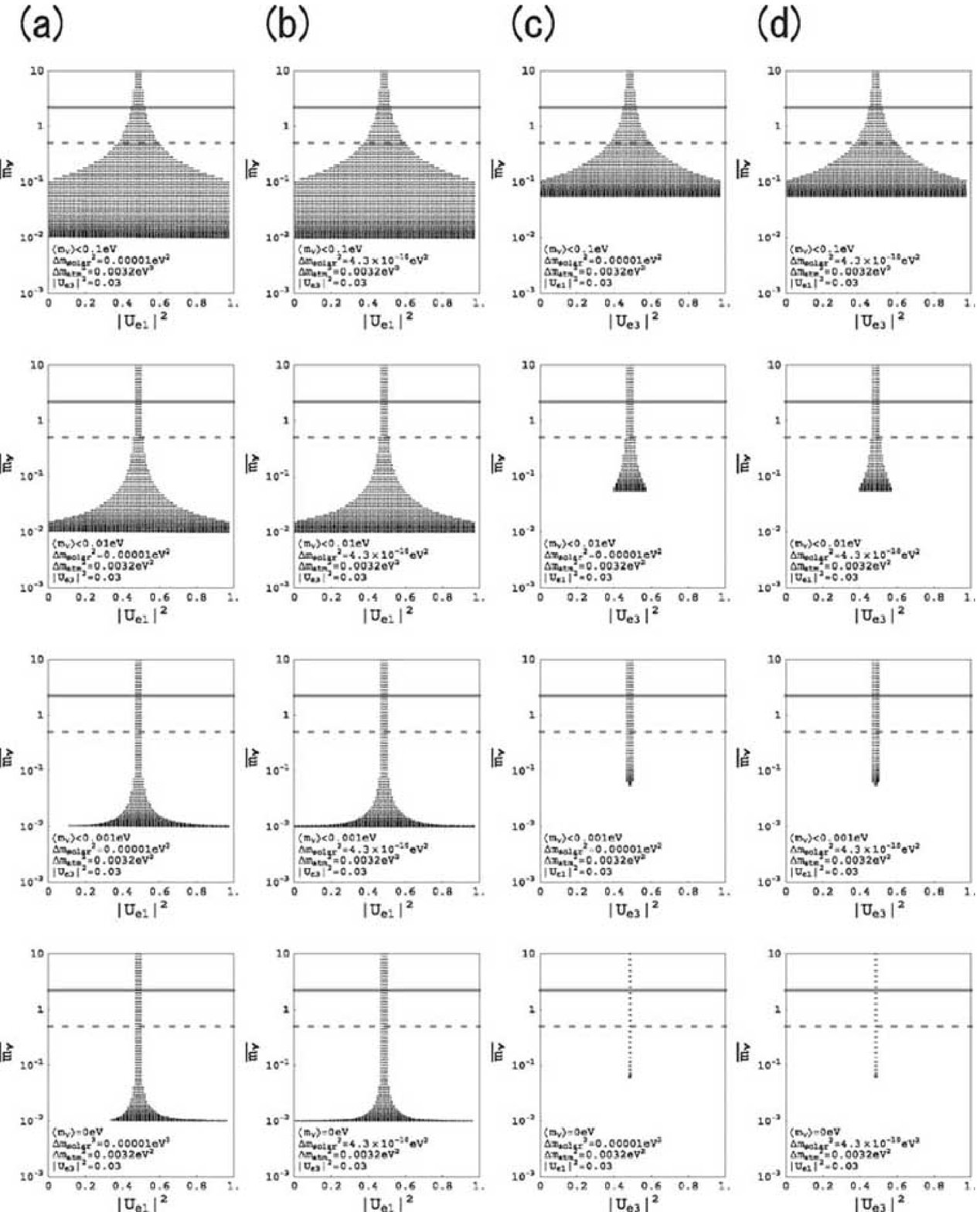}}
\caption{%
The allowed regions in the $\overline{m_{\nu}}$-$|U_{e1}|^2$ plane 
for fixed values of 
$\langle m_{\nu} \rangle_{max}$, $\Delta m_{solar}^2$, 
$\Delta m_{atm}^2$, and $|U_{e3}|^2$ 
in the normal mass hierarchy scenario ((a) and (b)), 
and in the $\overline{m_{\nu}}$-$|U_{e3}|^2$ plane for 
fixed values of $\langle m_{\nu} \rangle_{max}$, 
$\Delta m_{solar}^2$, $\Delta m_{atm}^2$, 
and $|U_{e1}|^2$ in the inverse mass hierarchy scenario ((c) and (d)). 
Here we choose $|U_{e3}|^2$ or 
$|U_{e1}|^2=0.03$, 
The allowed regions change very little when $|U_{e3}|^2$ or $|U_{e1}|^2<0.03$. 
The figures of the first and the second rows are for the MSW and Just So solutions 
for the solar neutrino problem respectively in the normal mass hierarchy scenario. 
The third and fourth ones are for the MSW and Just So solutions in the inverse 
mass hierarchy scenario. The first, second, third, and fourth columns are for 
$\langle m_{\nu} \rangle<0.1$ eV, $\langle m_{\nu} \rangle<0.01$ eV, 
$\langle m_{\nu} \rangle<0.001$ eV, 
and $\langle m_{\nu} \rangle=0$ eV, respectively. 
The gray solid (dotted) line shows 
the present bound \(\overline{m_\nu} < 2.2\)eV by Mainz
(the future bound  \(\overline{m_\nu} < 0.5\)eV) [2].}
\label{fig4}
\end{figure}
\begin{figure}[htbp]
\centerline{\epsfbox{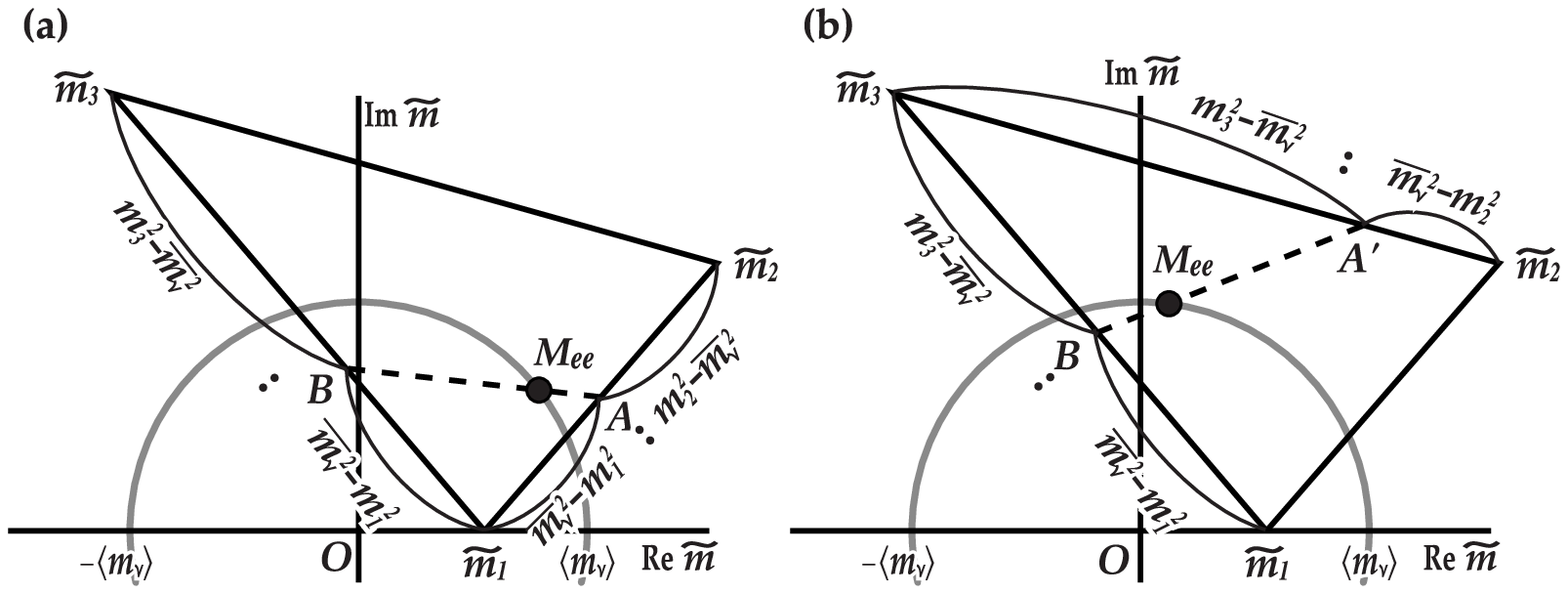}}
\caption{%
The allowed position of the complex mass $M_{ee}$ from $(\beta\beta)_{0\nu}$ 
and the single beta decay. The position of $M_{ee}$ is restricted to be on 
the line segment $AB$ in Fig.5(a) 
for the case $\overline{m_{\nu}}<m_2$ 
or $A^\prime B$ in Fig.5(b) for the case $\overline{m_{\nu}}>m_2$ 
with given definite 
values of $\overline{m_{\nu}}$, $m_i$, and one free parameter of mixing matrix 
element.}
\label{fig5}
\end{figure}
\begin{figure}[htbp]
\centerline{\epsfbox{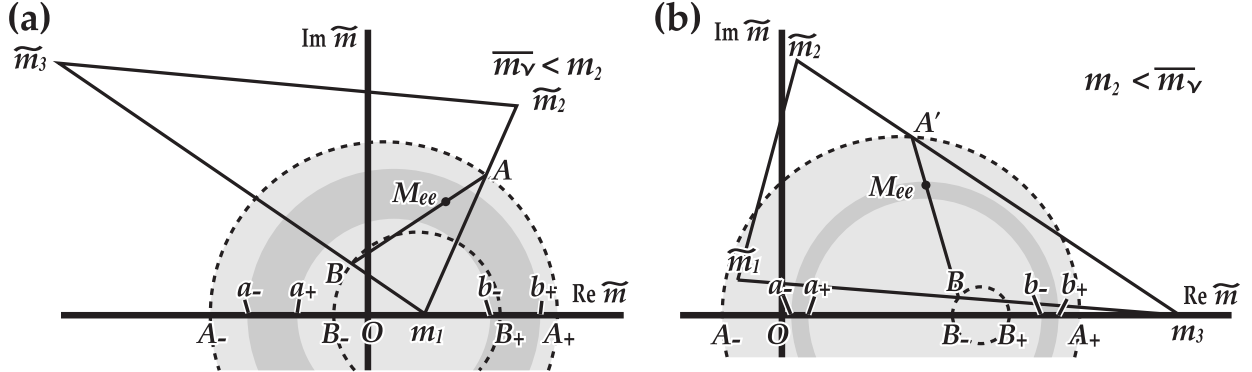}}
\caption{%
The allowed region of \(M_{ee}\) obtained by making the 
\(CP\) violating phases run freely for the cases in which
(a) \(\overline{m_{\nu}}<m_2\) and (b) \(\overline{m_{\nu}}>m_2\). 
The dark shaded region is the same as the shaded region in Fig.3,
in which we move the \(CP\) violating phases freely but fix the mixing angles.
The light shaded region is the allowed region for the case  
where mixing matrix elements are also free and
\(M_{ee}\) can move along 
the line segment \(AB\) or \(A^\prime B\), freely.
(i.e. All mixing matrix elements are free parameters.) 
In left-hand side figure (a), 
\(A_{-} \equiv\) \((m_1m_2-\overline{m_{\nu}}^2)\) \(/\) \((m_2-m_1)\),
\(A_{+} \equiv\) \((m_1m_2+\overline{m_{\nu}}^2)\) \(/\) \((m_2+m_1)\),
\(B_{-} \equiv\) \((m_1m_3-\overline{m_{\nu}}^2)\) \(/\) \((m_3-m_1)\) and
\(B_{+} \equiv\) \((m_1m_3+\overline{m_{\nu}}^2)\) \(/\) \((m_3+m_1)\).
The right-hand side figure (b) is obtained by exchanging the suffix 3 for 1
in (a).
Therefore, 
\(A_{-} \equiv\) \((-m_2m_3+\overline{m_{\nu}}^2)\) \(/\) \((m_3-m_2)\),
\(A_{+} \equiv\) \((m_2m_3+\overline{m_{\nu}}^2)\) \(/\) \((m_3+m_2)\),
\(B_{-} \equiv\) \((-m_1m_3+\overline{m_{\nu}}^2)\) \(/\) \((m_3-m_1)\) and 
\(B_{+} \equiv\) \((m_1m_3+\overline{m_{\nu}}^2)\) \(/\) \((m_3+m_1)\).
}

\label{fig6}
\end{figure}

(i) By changing the $CP$ violating phases freely we have discussed 
the consistency condition among the neutrinoless double beta decay and 
neutrino oscillations for the two hierarchy scenarios of neutrino masses 
(normal and inverse mass hierarchies). 
In this analysis we have used only observable neutrino mass parameters 
such as $\langle m_{\nu} \rangle$, $\overline{m_{\nu}}$, $\Delta m_{12}^2$, 
and $\Delta m_{23}^2$ 
instead of using $m_1$, $m_2$, and $m_3$ by incorporating the single beta decay also. 
It should be noted that $m_1$, $m_2$, and $m_3$ are related 
to the observable parameters as seen in Eqs.(3.6)-(3.8). 
Then, we find that SMA-MSW for 
the solar neutrino problem are disfavored for small \(\langle m_\nu \rangle\) 
($\leq 0.01$ eV)
in the inverse neutrino mass hierarchy scenario. 
This is the case even in the normal mass hierarchy scenario 
except for very restrictive value of $\overline{m_{\nu}} \sim 1\times 10^{-2}$ eV.  
In addition to this, 
we have also found that the neutrino mass observed in the single beta decay 
$\overline{m_{\nu}}$ 
is predicted to be larger than $O(10^{-2})$ eV
for the normal neutrino mass hierarchy scenario and than $O(10^{-1})$ eV for the
inverse neutrino mass hierarchy scenario. Those results are shown in Fig.\ref{fig4}.
\par
(ii) We have discussed the allowed region in the two neutrino masses plane, 
$\langle m_{\nu} \rangle$ and $\overline{m_{\nu}}$ which are observable 
in $(\beta\beta)_{0\nu}$ and single beta decay respectively. 
First we have obtained Eqs.(4.8) and (4.9) which are irrelevant to
the mixing angles (Fig.7).  Then more stringent constraints are obtained by 
incorporating the data of mixing angles (Fig.8).
\par
(iii)We have analytically obtained the constraints, 
Eqs.(5.1) and (5.2), 
on the Majorana $CP$ violating phases for typical limits of mixing angle with use of 
the observable neutrino masses ($\langle m_{\nu} \rangle$ and $\overline{m_{\nu}}$). 
The $m_1$, $m_2$, and $m_3$ 
in these constraints are described in terms of observables 
using Eqs.(\ref{eq1220-11})-(\ref{eq1220-12}). 
Then we have obtained the relations among the \(CP\) phases 
and $\overline{m_{\nu}}$ for several possible values of $\langle m_{\nu} \rangle$, 
$|U_{e2}|^2$ with fixed $\Delta m_{solar}^2$ 
and $\Delta m_{atm}^2$, and $|U_{e3}|^2$.
For the case of \(|U_{e1}|^2\simeq |U_{e1}|^2\simeq 1/2\), 
\(CP\) phases are severely constrained if $\langle m_\nu \rangle$ 
and $\overline{m_{\nu}}$ have nonzero values in the neighborhood 
of the present experimental upper limits (Fig.9 and Fig.10).\\

We are grateful to T. Ohshima for comments. 
The work of K.M. is supported by the JSPS Research Fellowship, No.10421.


\newpage
%

\begin{figure}[htbp]
\centerline{\epsfbox{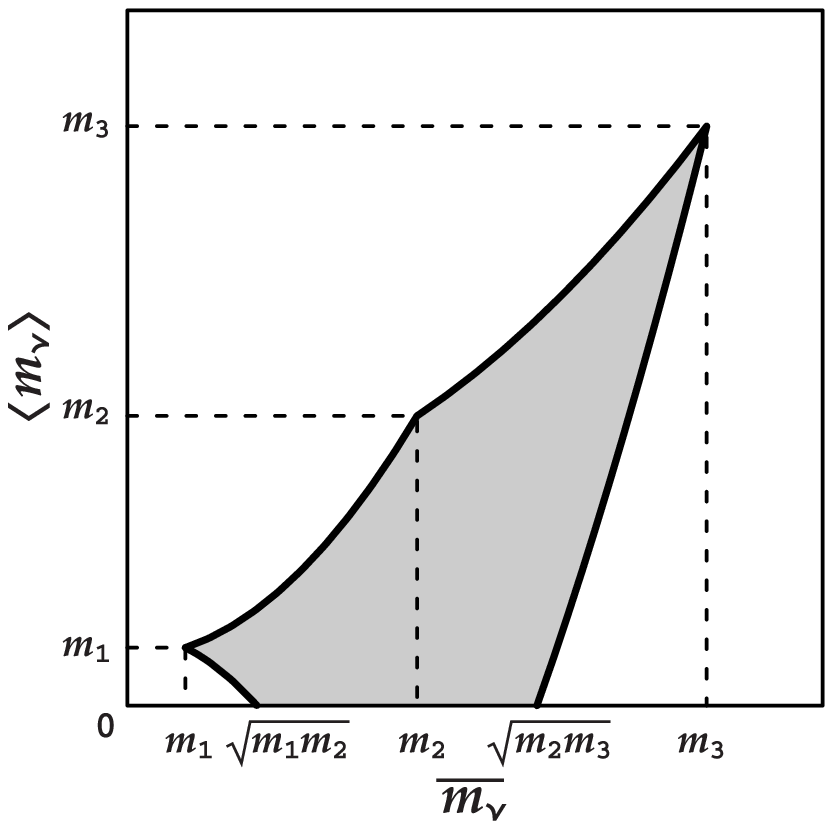}}
\caption{%
The allowed region in the $\langle m_{\nu} \rangle$-$\overline{m_\nu}$ plane obtained 
independently of the $CP$ violating phases. 
This constraint is obtained by considering Fig.6.
The boundary curves are given by Eqs. (4.8) and (4.9).}
\label{fig7}
\end{figure}

%

\begin{figure}[htbp]
\centerline{\epsfbox{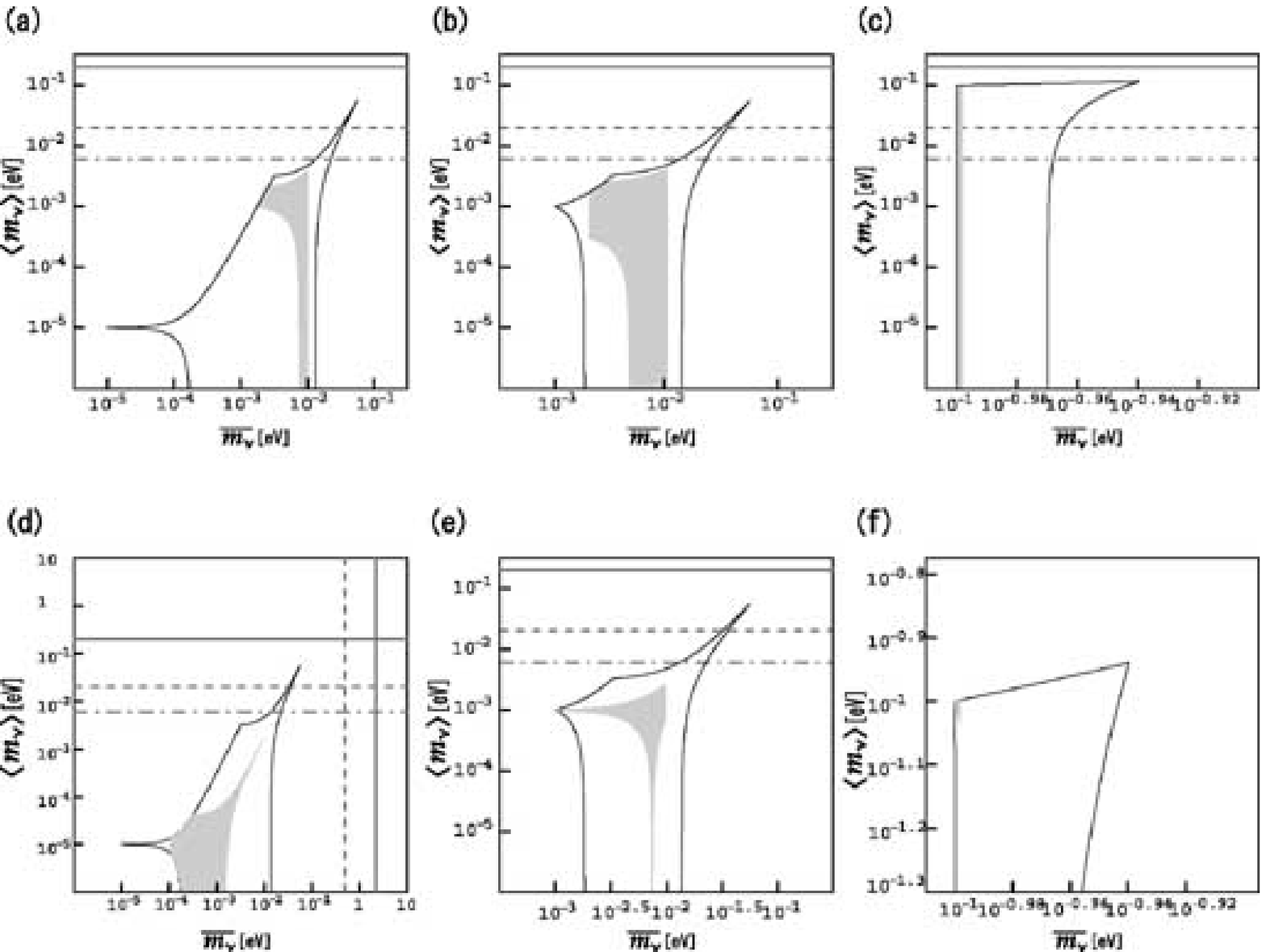}}
\caption{The allowed region 
in the \(\langle m_\nu\rangle\)-\(\overline{m_\nu}\) plane 
in the normal mass hierarchy obtained independently of the \(CP\) violating phases. 
In this figure the data of the mixing angles are incorporated:
$|U_{e3}|^2<0.03$ and  $0.3<|U_{e2}|^2<0.7$ for LMA-MSW (the upper panels) 
and $|U_{e3}|^2<0.03$, $1\times 10^{-3}<|U_{e2}|^2<1\times 10^{-2}$ 
for SMA-MSW (the lower panels).  
The first, second and third row panels correspond 
to the case $m_1=10^{-5}$eV, $10^{-3}$eV, $10^{-1}$eV, respectively. 
The black solid curve is the boundary of Fig.7.
The horizontal lines show the present bound 
\(\langle m_\nu \rangle < 0.2\) eV by Heidlberg-Moscow  (gray solid line),
the future bounds \(\langle m_\nu \rangle < 0.02\)  eV by GENIUS I (gray dotted line)  
and \(\langle m_\nu \rangle < 0.006\) eV by GENIUS II (gray dot-dashed line) [6].
The vertical lines show 
the present bound \(\overline{m_\nu} < 2.2\)eV by Mainz (gray dotted line)
and the future bound  \(\overline{m_\nu} < 0.5\)eV (gray dotted line)[2], 
which are on the right out of the panel except for the case (d). 
In the case of (f), 
the present limit of $\langle m_\nu\rangle$ 
is above and GENIUS I, II are below the panel. 
}
\label{fig9}
\end{figure}

\begin{figure}[htbp]
\centerline{\epsfbox{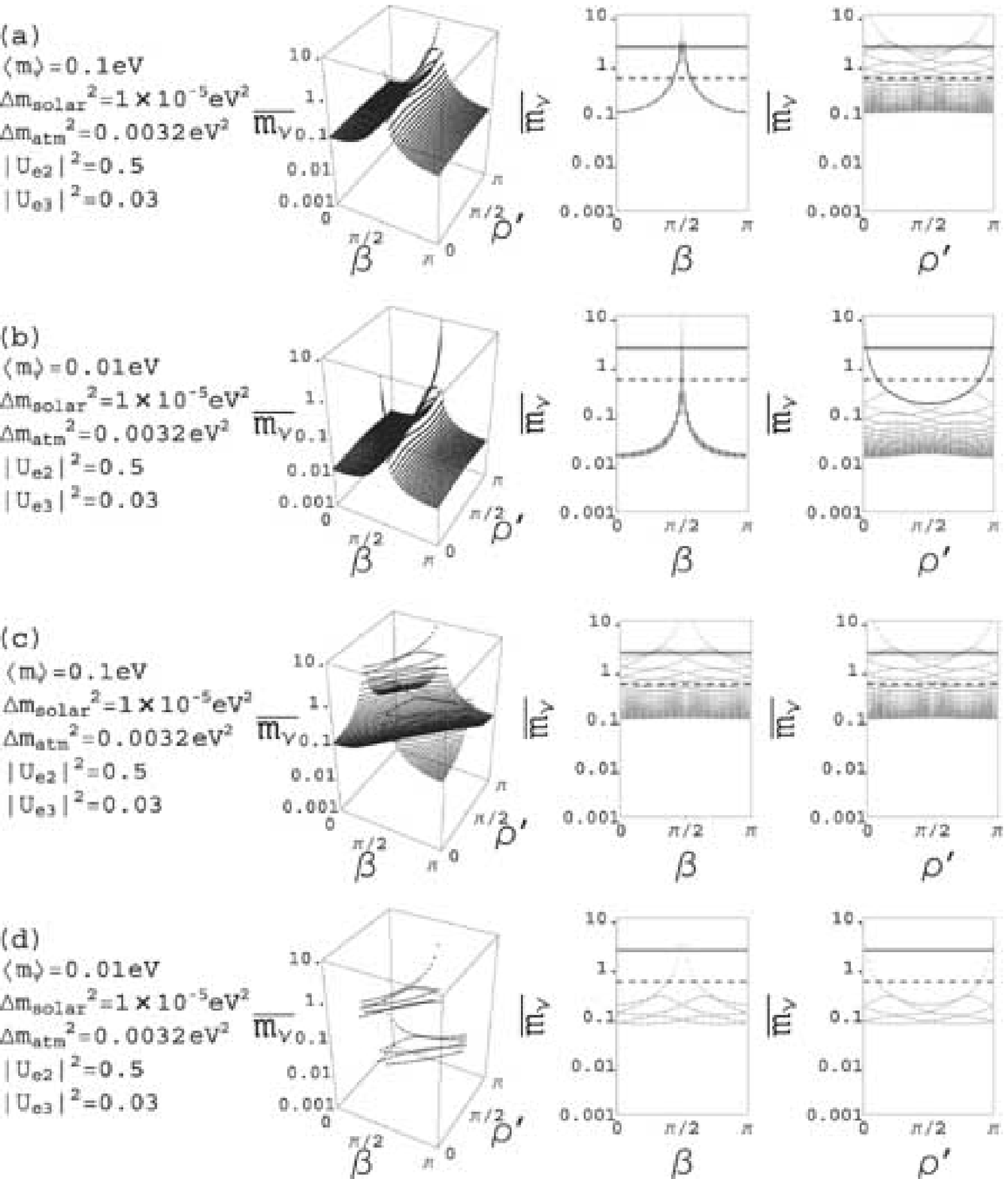}}
\caption{The allowed region among $\overline{m_{\nu}}$, $CP$ violating 
phases $\beta $, and $\rho '$ 
for the LMA-MSW solution for the solar neutrino problem in the normal 
((a) and (b)) and inverse ((c) and (d)) neutrino mass hierarchy.
We take the following values: 
$\langle m_{\nu} \rangle$ = 0.1 eV and 0.01 eV and the others are fixed as 
$|U_{e2}|^2=0.5,~|U_{e3}|^2=0.03,~\Delta m_{solar}^2=1\times 10^{-5}$eV$^2$ 
and $\Delta m_{atm}^2=3.2\times 10^{-3}$eV$^2$.
The gray solid (dotted) line shows 
the present bound \(\overline{m_\nu} < 2.2\)eV by Mainz
(the future bound  \(\overline{m_\nu} < 0.5\)eV) [2].
}
\label{fig10}
\end{figure}

\begin{figure}[htbp]
\centerline{\epsfbox{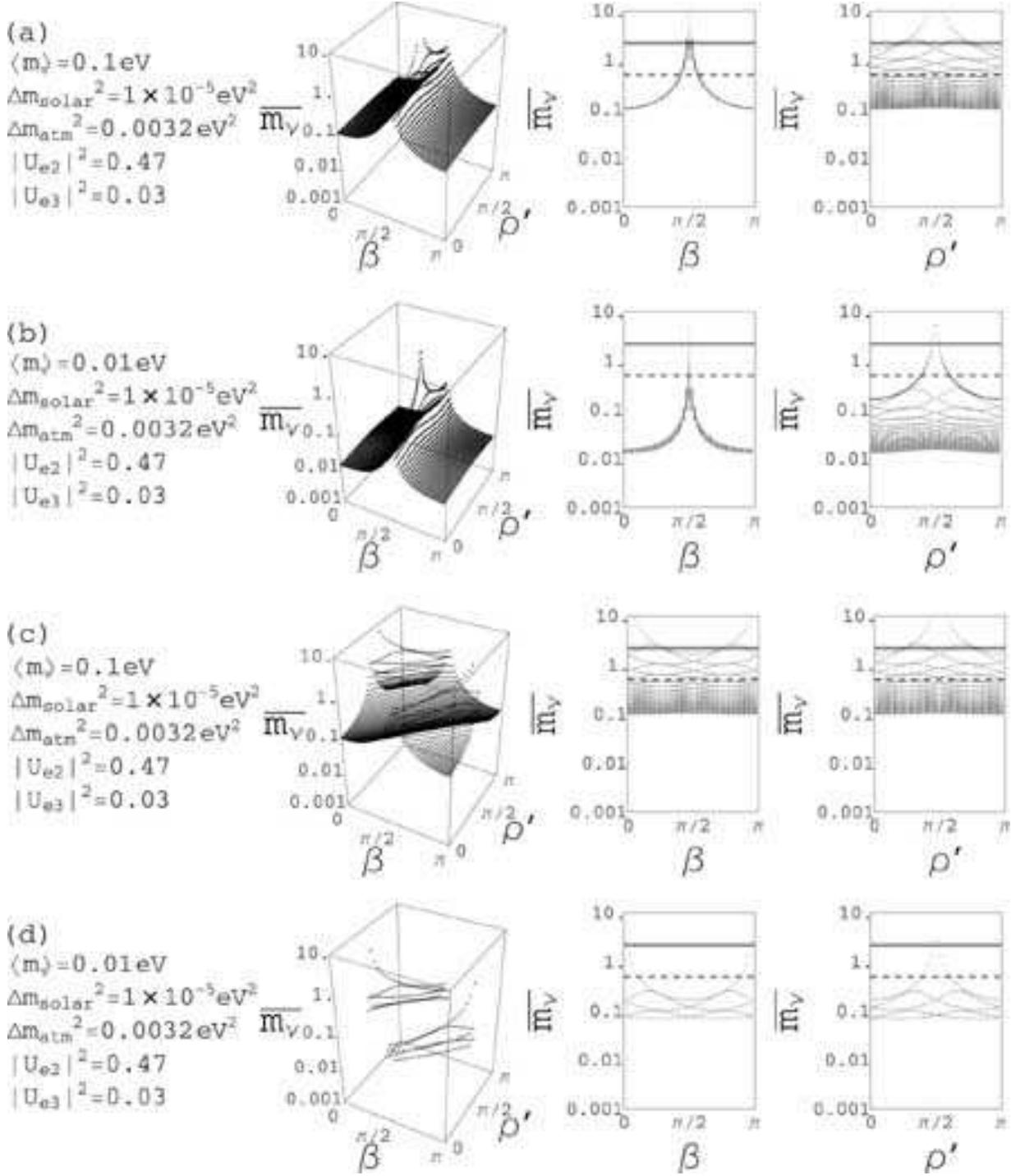}}
\caption{%
Same as Fig.9 except for \(|U_{e2}|^2=0.47\) in place of \(|U_{e2}|^2=0.5\).
}
\label{fig11}
\end{figure}

\end{document}